\documentclass[11pt,a4paper]{article}
\usepackage{amssymb} \usepackage{amsmath} \usepackage{graphicx}
\usepackage{slashed}
\usepackage{epsfig,latexsym,color,xcolor}
\usepackage{ulem}
\usepackage{amstext}    % defines the \text command, needed here
\usepackage{array} 
\usepackage{amsmath}
\begin{document}
%%%%%%%%%%%
\def\Giulia{\bf\color{red}}
%%%%%%%%%%%%%
\def\bef{\begin{figure}}
\def\eef{\end{figure}}
\newcommand{\ans}{ansatz }
\newcommand{\be}[1]{\begin{equation}\label{#1}}
\newcommand{\beq}{\begin{equation}}
\newcommand{\ee}{\end{equation}}
\newcommand{\beqn}[1]{\begin{eqnarray}\label{#1}}
\newcommand{\eeqn}{\end{eqnarray}}
\newcommand{\bd}{\begin{displaymath}}
\newcommand{\ed}{\end{displaymath}}
\newcommand{\mat}[4]{\left(\begin{array}{cc}{#1}&{#2}\\{#3}&{#4}
\end{array}\right)}
\newcommand{\matr}[9]{\left(\begin{array}{ccc}{#1}&{#2}&{#3}\\
{#4}&{#5}&{#6}\\{#7}&{#8}&{#9}\end{array}\right)}
\newcommand{\matrr}[6]{\left(\begin{array}{cc}{#1}&{#2}\\
{#3}&{#4}\\{#5}&{#6}\end{array}\right)}
\newcommand{\cvb}[3]{#1^{#2}_{#3}}
\def\lsim{\raise0.3ex\hbox{$\;<$\kern-0.75em\raise-1.1ex
e\hbox{$\sim\;$}}}
\def\gsim{\raise0.3ex\hbox{$\;>$\kern-0.75em\raise-1.1ex
\hbox{$\sim\;$}}}
\def\abs#1{\left| #1\right|}
\def\simlt{\mathrel{\lower2.5pt\vbox{\lineskip=0pt\baselineskip=0pt
           \hbox{$<$}\hbox{$\sim$}}}}
\def\simgt{\mathrel{\lower2.5pt\vbox{\lineskip=0pt\baselineskip=0pt
           \hbox{$>$}\hbox{$\sim$}}}}
\def\unity{{\hbox{1\kern-.8mm l}}}
\newcommand{\eps}{\varepsilon}
\def\ep{\epsilon}
\def\ga{\gamma}
\def\Ga{\Gamma}
\def\om{\omega}
\def\omp{{\omega^\prime}}
\def\Om{\Omega}
\def\la{\lambda}
\def\La{\Lambda}
\def\al{\alpha}
\def\beq{\begin{equation}}
\def\eeq{\end{equation}}
\newcommand{\ov}{\overline}
\renewcommand{\to}{\rightarrow}
\renewcommand{\vec}[1]{\mathbf{#1}}
\newcommand{\vect}[1]{\mbox{\boldmath$#1$}}
\def\tm{{\widetilde{m}}}
\def\mcirc{{\stackrel{o}{m}}}
\newcommand{\Dm}{\Delta m}
\newcommand{\dm}{\varepsilon}
\newcommand{\tanb}{\tan\beta}
\newcommand{\nbar}{\tilde{n}}
\newcommand\PM[1]{\begin{pmatrix}#1\end{pmatrix}}
\newcommand{\up}{\uparrow}
\newcommand{\down}{\downarrow}
\newcommand{\refs}[2]{eqs.~(\ref{#1})-(\ref{#2})}
\def\omE{\omega_{\rm Ter}}
\newcommand{\eqn}[1]{eq.~(\ref{#1})}
%
%%%%%%%%%%     mauri    %%%%%%%%%%%%%%%%%%%%%%%%%%%%%%%%%

\newcommand{\DSUSY}{{SUSY \hspace{-9.4pt} \slash}\;}
\newcommand{\DCP}{{CP \hspace{-7.4pt} \slash}\;}
\newcommand{\mc}{\mathcal}
\newcommand{\gr}{\mathbf}
\renewcommand{\to}{\rightarrow}
\newcommand{\gtc}{\mathfrak}
\newcommand{\wh}{\widehat}
\newcommand{\br}{\langle}
\newcommand{\kt}{\rangle}

%%%%%%%%%%%%%%%%%%%%%%%%%%%%%%%%%%%%%%%%%%%%%%%%%%%%%%%%%%

% barbara Ricci  %definizione di minore e maggiore simile
\def\lsim{\mathrel{\mathop  {\hbox{\lower0.5ex\hbox{$\sim$}
\kern-0.8em\lower-0.7ex\hbox{$<$}}}}}
\def\gsim{\mathrel{\mathop  {\hbox{\lower0.5ex\hbox{$\sim$}
\kern-0.8em\lower-0.7ex\hbox{$>$}}}}}
%%%%%%%%%%%%%%%%%%%%%%%%%%%%%%%%%%

\def\nn{\\  \nonumber}
\def\de{\partial}
\def\brf{{\mathbf f}}
\def\bbf{\bar{\bf f}}
\def\bF{{\bf F}}
\def\bbF{\bar{\bf F}}
\def\bA{{\mathbf A}}
\def\bB{{\mathbf B}}
\def\bG{{\mathbf G}}
\def\bI{{\mathbf I}}
\def\bM{{\mathbf M}}
\def\bY{{\mathbf Y}}
\def\bX{{\mathbf X}}
\def\bS{{\mathbf S}}
\def\bb{{\mathbf b}}
\def\bh{{\mathbf h}}
\def\bg{{\mathbf g}}
\def\bla{{\mathbf \la}}
\def\bmu{\mathbf m }
\def\by{{\mathbf y}}
\def\bmu{\mbox{\boldmath $\mu$} }
\def\bsig{\mbox{\boldmath $\sigma$} }
\def\bunity{{\mathbf 1}}
\def\cA{{\cal A}}
\def\cB{{\cal B}}
\def\cC{{\cal C}}
\def\cD{{\cal D}}
\def\cF{{\cal F}}
\def\cG{{\cal G}}
\def\cH{{\cal H}}
\def\cI{{\cal I}}
\def\cL{{\cal L}}
\def\cN{{\cal N}}
\def\cM{{\cal M}}
\def\cO{{\cal O}}
\def\cR{{\cal R}}
\def\cS{{\cal S}}
\def\cT{{\cal T}}
\def\eV{{\rm eV}}

\numberwithin{equation}{section}
%\today

%\begin{flushright}
%CERN-TH-2016-232\\
%\end{flushright}
\vspace{6mm}

\large
 \begin{center}
 %{\Large \bf Gravitational waves signals from Dark Exotic Stars : 
% mirror neutron stars merging }
 {\Large \bf Testing merging of Dark Exotic Stars from 
 
 Gravitational Waves in the Multi-messanger approach.}

 \end{center}

 \vspace{0.1cm}

% \vspace{0.1cm}
% \begin{center}
%{\large Andrea Addazi}\footnote{E-mail: \,  andrea.addazi@infn.lngs.it} \\

%{\it \it Dipartimento di Fisica,
% Universit\`a di L'Aquila, 67010 Coppito, AQ \\
%LNGS, Laboratori Nazionali del Gran Sasso, 67010 Assergi AQ, Italy}
%\end{center}

\begin{center}
{\large Andrea Addazi \& Antonino Marciano}\footnote{E-mail: \, andrea.addazi@qq.com, marciano@fudan.edu.cn}
\\
{\it Department of Physics \& Center for Field Theory and Particle Physics, Fudan University, 200433 Shanghai, China}
\end{center}

\vspace{1cm}
\begin{abstract}
\large
\noindent 
We discuss possible implications of the recent detection by the LIGO and VIRGO collaboration of the gravitational-wave event GW170817, the signal of which is consistent with predictions in general relativity on the merging of neutron stars. A near-simultaneous and spatially correlated observation of a gamma-ray burst, the GRB 170817A signal, was achieved independently by the Fermi Gamma-ray Burst Monitor, and by the Anti-coincidence Shield for the Spectrometer of the International Gamma-Ray Astrophysics Laboratory. Motivated by this near temporal and spatial concomitance of events, which can occur by chance only with the probability $5.0 \, \times \, 10^{-8}$, we speculate on the possibility that new dark stars signals could be detected from the LIGO/VIRGO detectors. This proposal, which aims at providing a test for some models of dark matter, relies on the recent achievement of detecting, for visible ordinary matter, the merging of neutron stars both in the gravitational and the electromagnetic channel. A lack of correlation between the two expected signals would suggest a deviation from the properties of ordinary matter. Specifically, we focus on models of invisible dark matter, and in particular we study the case of mirror dark matter, within the framework of which a large amount of mirror neutron stars are naturally envisaged to occupy our dark matter halo.  The observation of an electromagnetically hidden event inside the dark matter halo of our galaxy should provide a hint of new physics. There would be indeed no satisfactorily complete explanation accounting for the lack of electromagnetic signal, if only standard neutron star merging were considered to describe events that happen so close to us.

\end{abstract}
%\newpage

\baselineskip = 20pt

\section{Introduction}

\begin{figure}[t]
\centerline{ \includegraphics [height=6cm,width=0.4\columnwidth]{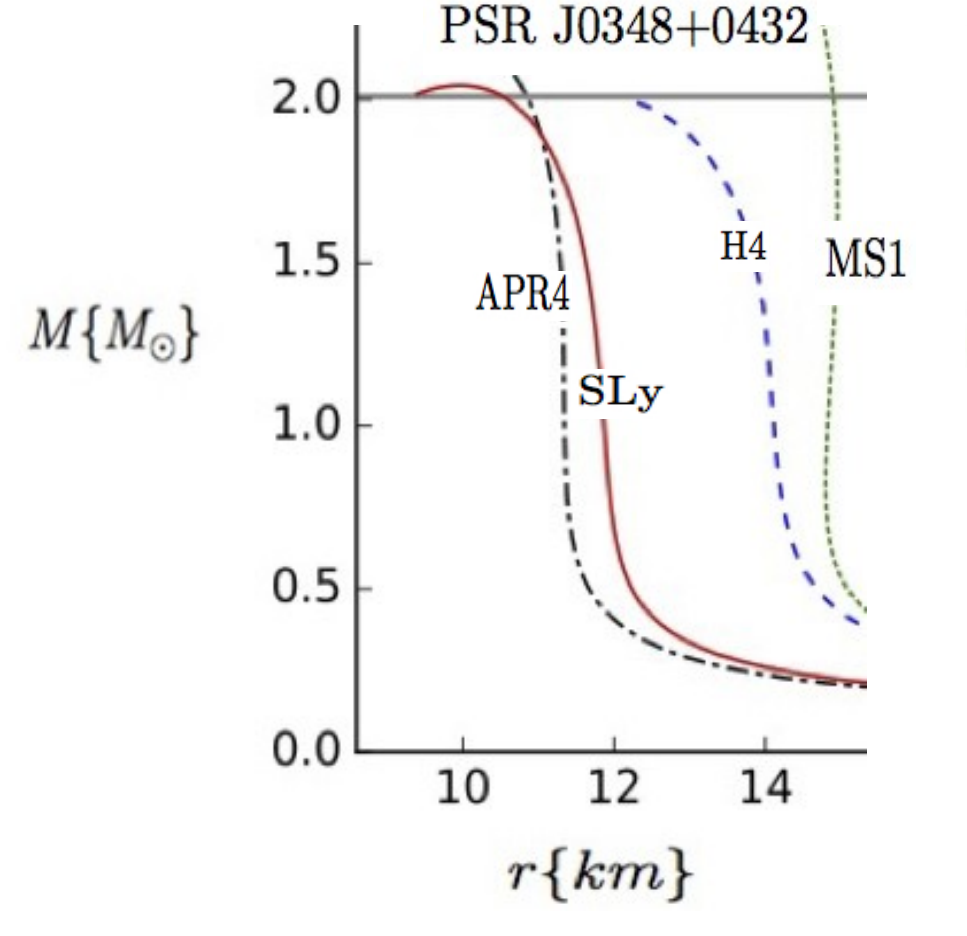}}
\vspace*{-1ex}
\caption{Realistic Mirror Model Equation of States are displayed, plotting the mass profile against the neutron star radius. Four EoS models are considered: the APR4 \cite{T142} (black), the SLy \cite{T140} (red), the H4 \cite{T101} (blu) and the MS1 (red) \cite{T143} models. We also report the experimental mass bound on the ordinary neutron stars, recovered from the PSRJ0543+1559 system, the most massive neutron star ever observed. }
\label{plot}   % \ref{plot}
\end{figure}

\begin{figure}[t]
\centerline{ \includegraphics [height=5.5cm,width=0.3\columnwidth]{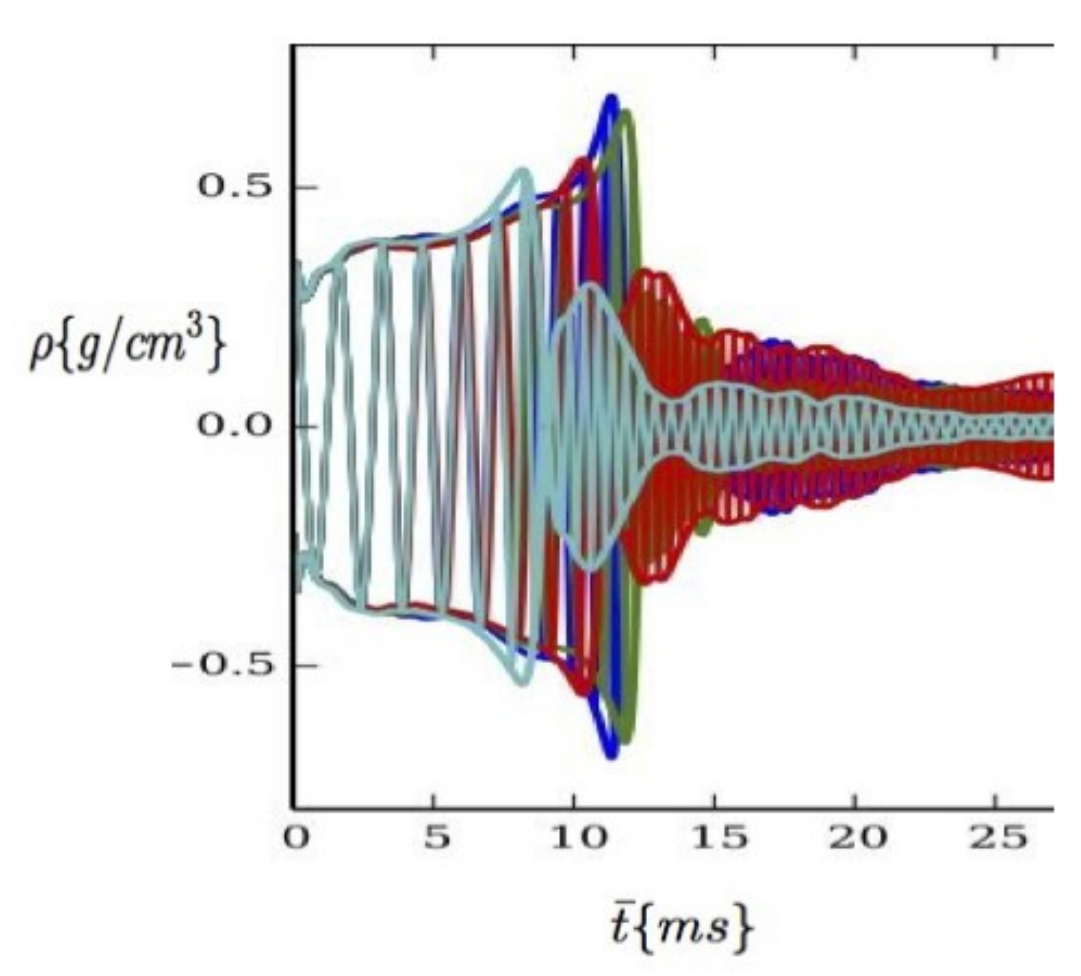}}
\vspace*{-1ex}
\caption{The energy-density evolution (related to the gravitational wave strain $rh_{22}^{+}$) is displayed as a function of time. We consider the case in which all the Neutron stars have the same mass, and an initial realistic inter-distance between the two Mirror neutron stars (we sets $40\, {\rm km}$) is taken into account. The different signals predicted from the various EoS models, already displayed in Fig.1, concern the APR4 \cite{T142} (blue), the SLy \cite{T140} (green), the H4 \cite{T101} (red) and the MS1 (light blue) \cite{T143} models.}
\label{plot}   % \ref{plot}
\end{figure}

\noindent 
The recent discovery of the gravitational-wave event GW170817 by the LIGO and VIRGO collaboration was the first observation of a gravitational wave combined with a gamma rays signal. The latter was detected thanks to the FERMI and the INTEGRAL collaborations, and to other constraints from multi-messengers observations. This concomitance of signals opens a clear pathway to the development of a new era of observations in astrophysics and cosmology \cite{Monitor:2017mdv,Abbott:2017ntl,GBM:2017lvd}.  At the level of the knowledge of the Universe's microphysics, data are compatible with the signal expected in general relativity (GR) from the merging of neutron stars, thus confirming predictions of the Einsteinian theory of gravity at an high accuracy level.

Following a line of thought that hinges on the current availability of data sets in both the gravitational and the electromagnetic channels, at least for a class of phenomena such as neutron stars merging, we may envisage the intriguing possibility to test dark exotic stars merging scenarios. The underlying idea consists on a focus on the multi-messenger approach in order to distinguish the observation of ordinary matter stars --- predicted from General Relativity and the Standard Model of particle physics --- from the observation of new exotic stars. These latter are predicted by the extensions of the Standard Model that allow for {\it new dark gauge sectors} and dark matter particles.\\

Contrary to standard WIMP candidates of dark matter, there are many possible extensions of the Standard Model that explain dark matter while predicting a richer astrophysical complexity than WIMP candidates, eventuallty hidden in the dark halo. The paradigm that the Standard Model could be extended with a hidden gauge sector, namely $G'$, as $SU(3)\times SU(2) \times U(1) \times G'$ is an old-standing idea and it was explored accounting for several different choices of $G'$. Within this scenario, standard model particles are assumed to be sterile with respect to the dark gauge group. 

The first proposal of hidden sector that was ever considered in the literature is the Mirror Dark Matter (MDM) model, suggested by {\it T.D.~Lee} and {\it C.N.~Yang} in their Noble prize paper Ref.~\cite{LY}. The phenomenological implications of the Mirror World were first analyzed by {\it Kobzarev, Okun} and {\it Pomeranchuk} in Ref.~\cite{3}, and then hitherto elaborated in Refs.~\cite{4,5,6,KT,7,8,9,10a,10b,11}. After these seminal papers, many astrophysical and phenomenological aspects were investigated, independently by {\it Z.~Berezhiani} {\it et al} \cite{Berezhiani:1995yi,Berezhiani:1995am,Berezhiani:2000gw,Berezhiani:2003wj,Berezhiani:2005hv,Berezhiani:2005ek,Berezhiani:2003xm,Bento:2001rc,Berezhiani:1996sz,Berezhiani:2008bc,Berezhiani:2008gi,Berezhiani:2006je,Berezhiani:2005vv} and by {\it R.~Foot} {\it et al}  \cite{Foot:1995pa,Foot:1999ph,Foot:2003iv,Foot:2004wz,Ciarcelluti:2008qk,Foot:2008nw,Ciarcelluti:2010dm,Foot:2013nea,Foot:2014mia,Clarke:2015gqw,Foot:2016wvj} --- see also Ref. \cite{Okun:2006eb} for a review by {\it L.B.~Okun} on this topic. \\
  
The Mirror world is a hidden replica of the Standard Model gauge group $G'_{SM}=SU'(3)\times SU'(2)\times U'(1)$ that restores the left-right symmetry violated in the Standard Model by the weak interactions. In other words, the Mirror World is a parity completion of the Standard Model as a $G_{SM}\times G_{SM}'$ gauge theory. The complete theory $G_{SM}\times G_{SM}'$ is invariant under Mirror discrete transformations as $G_{SM}\longleftrightarrow G'_{SM}$. If Mirror parity is an exact discrete symmetry, then it must relate every ordinary particles, including the electron $e$, the proton $p$, the neutron $n$, the photon $\gamma$, the neutrinos $\nu$ etc. with their mirror twins $e'$, $p'$, $\gamma'$, $\nu'$ etc. Mirror partners are sterile to ordinary strong, weak and electromagnetic interactions, but in turn undergo their own hidden gauge interactions, belonging to the $SU'(3)\times SU'(2)\times U'(1)$ hidden sector, with exactly the same Ordinary Standard Model couplings. Thus, no new extra coupling parameters describing the Mirror physics are introduced. Ordinary and Mirror twin particles have the same mass
and the same microphysics rules, from high energy particle physics to atomic and nuclei physics. \\

Minimal version of MDM have been also considered in the literature, especially in light of their cosmological applications. These models usually double a subgroup of the Standard Model gauge group --- see {\it e.g.} Ref.~\cite{RM1, RM2, RM3, Alexander:2016xbm, Addazi:2016nok}, and do not necessarily preserve the dark visible symmetry at the level of the coupling constants of the theory. This might then allow for a resolution of both the dark energy problem and the dark matter problem, at the same time, by allowing for an imbalance between the energy scales of the dark matter and quintessence fields. We are not anyway biased in proposing this working direction. Within this analysis we take a minimalistic approach, so to be as much as possible general about our assumptions and encode the most generic cases.\\

The plan of the paper is the following. In Sec.~\ref{mico} we address the cosmological constraints of MDM models, extending the analysis on minimal models making use only of subgroups of the Standard Model of particle physics. In Sec.~\ref{DH} we review structure formation within the framework of MDM models, and illustrate the complexity of the dark matter scenario and its constraints for MDM models. In Sec.~\ref{GWS} we review theoretical predictions from General Relativity on the gravitational waves signal. Finally, in Sec.~\ref{co} we spell out conclusions and specify future outlooks for research in this direction.

\section{Mirror Cosmology} \label{mico}

\noindent 
We start reviewing the peculiar cosmological features that must be fulfilled by MDM models, in order to become valid models of dark matter. We then generalize the discussion so to encode minimal and partial scenarios of MDM. \\

Although the Ordinary and the Mirror Sectors share the same microphysics, a reliable dark matter model must posses a specific cosmological evolution consistent with experimental constraints. For instance, if we naively assume that $\Omega_{B}'=\Omega_{B}$, in which $\Omega_{B}$ and $\Omega_{B}'$ are the relative amounts of Ordinary and Mirror Baryons, then the Mirror Matter cannot provide for all the amount of Cold Dark Matter $\Omega_{DM}\simeq 5\Omega_{B}$. On the other hand $T'=T$, {\it i.e.} an equal temperature between the two sectors, would be in inconsistent with limits from Big Bang Nucleosynthesis (BBN) on the amount of relativistic degrees of freedom. Extra Mirror neutrinos must contribute as $\Delta N_{eff}=6.15$, which is inconsistent with the experimental bound on it --- close to $\Delta N_{eff}^{exp}\simeq 0.5$. \\

\noindent 
To avoid encountering this kind of shortcomings, we must assume that:
\begin{enumerate}

\item 
after the inflationary epoch, the two sectors were re-heated with different temperatures;

\item
the Ordinary and Mirror Worlds are very weakly interacting, in such a way that both the systems evolve adiabatically, without undergoing into thermal equilibrium. 

\end{enumerate}

\noindent 
Under these two hypotheses, the Mirror matter becomes a viable candidate of collisional and dissipative cold dark matter. \\

%--Comments on the Mirror BBN

Furthermore, BBN in the Mirror World will happen in a cosmological time antecedent to the time of BBN in the Ordinary World. The condition $T'\!<\!\!<\!T$ --- here $T'$ and $T$ are respectively the Mirror and the Ordinary temperatures --- implies that a larger amount of baryon asymmetry is produced in the Mirror sector, {\it i.e.} $\eta_{B}'>\eta_{B}\simeq (2\div 6)\times 10^{-10}$ \cite{Berezhiani:2000gw}.  As shown in Ref.~\cite{Berezhiani:2000gw}, as a consequence of the temperature asymmetry between the two sectors, the produced He-4 mass fraction is considerably larger than in the ordinary sector, since $Y_{4}'\simeq 0.4 \div 0.8$ against $Y_{4}\simeq 0.24$. Thus He-4 constitutes the largest amount of MDM during recombination epoch. \\

We emphasize that the aforementioned issue of recovering extra relativistic degrees of freedom is solved when $T'\!<\!\!<\!T$. Basically, the number of degrees of freedom $g_{*}$ gets a contribution from the Mirror sector $(e',\nu',\gamma'_{e,\mu,\tau})$, which is suppressed with the ratio $(T'/T)^{4}$ \cite{Berezhiani:2000gw}. Consequently, since the 
extra degrees of freedom are found to be  
$$\Delta g_{*}=g_{*}(1+x^{4})-10.75=1.75 \, \Delta N_{\nu}\, , $$
we recover the bound 
$$\Delta N_{\nu}=6.14\, x^{4}\rightarrow x<0.64\, ,$$
which is highly compatible with the``Mirror Matter=Dark Matter" bound, {\it i.e.} 
$$\Omega_{B'}/\Omega_{B}\simeq 5\,.$$ 
This opens the pathway to the possibility of reconciling neutrinos flux anomalies in short baseline, understandable in terms of extra sterile neutrinos, with the BBN bounds \cite{Berezhiani:1995yi,Foot:1995pa}.\\

Minimal dark models that double only subgroups of the Standard Model gauge group may be taken into account in stead, and finally encoded into this proposal on correlated GW and GRB detection, making more general our discussion. For instance, discrete symmetries are present in some extensions of the Standard Model, in which we encounter vector-like couplings of the electroweak SU(2) subgroup to an ``hidden'' fermionic sector undergoing confinement by strong forces. This was for instance considered in Ref.~\cite{RM3}, in which the hidden fermionic sector is shown to lead to the formation of isotriplet.

On the other hand, doubling directly the QCD sector as in \cite{Alexander:2016xbm, Addazi:2016nok} without taking into account any discrete symmetry that involves the coupling constants of the visible and invisible sectors, can lead to scenarios in which the resolution of the dark energy paradigm entails an usual path toward the explanation of dark matter. For instance, fixing the pion decay constant to be $f_D\simeq$ meV, in order to reproduce constraints on the current accelerated expansion of the Universe, induces to set up a scale for invisible dark matter that predicts a halo composed of very compact dark neutron stars, strange/quark stars and black holes, the masses of which are $M_{\rm MACHO}< 10^{-7} M_\odot$.  \\

To make more focused our analysis, we will no further refer to this scenario here, and leave the reader to the relevant references.

\section{Structure Formation and Dark Halo} \label{DH}

\noindent
We focus in this section on the dark matter structure formation, specifying how the dark halo densities can be reproduced in MDM models. \\
  
As we mentioned above, a dark astrophysical complexity can be envisaged: nothing prevents the formation of galaxy structures and stars in both the extended and minimalistic Mirror sectors. Though, the dark stars formation will be radically different than the ordinary one, being highly affected from the initial nuclei composition arising from BBN. The problem of mirror stars formation was studied in Ref.~\cite{Berezhiani:2005vv}, in which using numerical methods for stars evolution and explosion time, the evolution of stars with large helium abundances (Y $= 0.30-0.80$) and a range of masses $0.5\div 10$ $ M_\odot $ was analyzed. The authors of Ref.~\cite{Berezhiani:2005vv} found that Helium dominated mirror stars have a much faster evolutionary time than ordinary stars with same masses. Since the abundance of Helium is higher in the Mirror sector, Mirror stars evolve faster as compared to ordinary ones. This means that they will explode earlier (as type II Supernovae) injecting Mirror nuclei in the Mirror interstellar medium. In turn, this will change the second generation stars formation processes, as well as the abundance of neutron stars, black holes, white and brown dwarfs. \\

The dynamical evolution of the dark halos emerging from this picture is highly complicated and unpredictable, since much more complicated that in the WIMP-like case. The problem of the Mirror neutron stars distributions in the Milky Way dark halo, as well as in other galaxies and clusters, is affected by a huge astrophysical uncertainties. Further considerations and remarks on the Mirror structure formation 
problem are more extensively discussed in the Appendix.

\section{Gravitational waves signals} \label{GWS}
\noindent 
In this section we show the gravitational waves signals, as expected from Mirror Neutron stars merging processes. The Equivalence Principle at the basis of General Relativity entails for Mirror twin objects exactly the same features as for the (exhaustively studied) Ordinary twin systems. Our task then trivializes into a simple review of results widely known in the literature.\\ 

Being more specific, we may model Mirror Neutron Stars equations as for the case of ordinary Neutron Stars. We then consider four realistic EoS models, {\it i.e.} the APR4 \cite{T142}, the SLy \cite{T140}, the H4 \cite{T101} and the MS1 \cite{T143} models. In Fig.1, we show the emitted energy-density in the gravitational waves channel, on the right side, in order to allow a comparison of results with the EoS displayed on the left side. In such a demonstrative case, we have assumed the mass of the two Neutron Stars to be equal to each other\footnote{Our results can be reproduced using open sources codes. In particular, we used {\it The Einstein Toolkit} in order to simulate the numerical evolution of the Hydrodynamical Equations in General Relativity \cite{T54}. We also consider the {\it LORENE} library \cite{T52,T53} for tabulated EoS.}. Our numerical results are in agreement with the large numerical GR literature dedicated on Neutron Stars Merging --- see {\it e.g.} \cite{Buonanno:1998gg,Damour:2009wj,Hinderer:2016eia,Steinhoff:2016rfi,Hotokezaka:2015xka,Hotokezaka:2016bzh,DePietri:2015lya,Maione:2016zqz,Feo:2016cbs,Maione:2017aux,Drago:2017bnf,Alford:2017rxf,Bovard:2017dfh,Hanauske:2016gia,Paschalidis:2016agf,Baiotti:2016qnr,Foucart:2015gaa,Dietrich:2017feu,R1,R2,R3,R4,R5,R6,R7,R8,R9,R10,R11}.\\
 
We may ask how many events for LIGO/VIRGO we should expect according to Mirror Dark Matter predictions, and where these should take place in or outside the galaxy. Unfortunately Mirror Dark Matter cannot provide a sharp answer to these questions, because of the aforementioned Mirror astrophysical complexity. Indicatively, if the Mirror Dark halo is distributed as a in the Galaxy Disk model proposed by Silk \cite{Silk}, then most of the galactic events are expected to be localized on the Milky Way (MW) Disk. On the other hand, within the case of the MW Elliptical Mirror halo, some events are also expected outside the Disk, where the gravitational waves noise is also much less present than in the Disk. Finally, most part of the events are expected to point toward the Galactic Bulge, since the Dark Matter halo has an over-density distribution in the center. Other events are expected outside our galaxy, in the proximity of Dark Matter halos, {\it i.e.} close to spirals, elliptical and irregular galaxies, clusters and so forth. 

Within the next subsections, we will further discuss technical details of our numerical analysis. In particular, our simulations take into account three main steps: i) the numerical simulations of the Einstein's and the hydrodynamical equations inside the neutron stars; ii) the numerical computation of the gravitational wave signal; iii) the modeling of the equation of state in the Mirror neutron star.

%\subsection{Technical numerical aspects of the analysis}

\subsection{Numerical solutions of Einstein's hydrodynamics equations }

\noindent
The numerical simulations of the Einstein's equations of fluids start from conveniently re-writing the $3+1$ covariant equations, separating the time coordinate from the spatial directions and foliating the space-time 
with space-like Cauchy surfaces, as usually done in the Arnowitt-Deser-Misner formulation (ADM)
\cite{203}.

A reformulation which is slightly different from ADM one is the so called {\bf Z4} formulation, which we have used in our numerical simulations, which can be derived from a Palatini-like variational principle \cite{219}
\be{LLL}
L=g^{\mu}(R_{\mu\nu}+2\nabla_{\mu}Z_{\nu})+\lambda(Z_{\mu})\,,
\ee
with $Z_{\mu}$ extra auxiliary vector field, with an algebraic constraint imposed from the Lagrangian multiplier $\lambda$: $Z_{\mu}=0$. From Eq.~\ref{LLL} one can easily derive the Equation of fields 
\be{RRR}
R_{\mu\nu}+Z_{\mu\nu}+k_{1}[n_{\mu}Z_{\nu}+n_{\nu}Z_{\mu}-(1+k_{2})g_{\mu\nu}n_{\rho}Z^{\rho}]
=8\pi\left(T_{\mu \nu}-\frac{1}{2}T\, g_{\mu\nu}\right)\, ,
\ee
plus the constraint $Z_\mu=0$, having also introduced 
\be{Zmunu}
Z_{\mu\nu}=\nabla_{\mu}Z_{\nu}+\nabla_{\nu}Z_{\mu}\,,
\ee
and with $k_{i}$ arbitrary numerical constants that can be conveniently selected to acquire any desired numerical value. 

By means of a conformal decomposition, and by separating the trace from the traceless components, after some tedious but straightforward algebra, one can obtain the following set of equations \cite{219}:
\be{eq1}
(\partial_{t}-\beta^{j}\partial_{j})K=-\gamma^{ij}\tilde{D}_{i}\tilde{D}_{j}\alpha+\alpha(R+K^{2}+2D_{i}Z^{i}-2\Theta K)-3\alpha k_{1}(1+k_{2})\Theta+4\pi \alpha(- 3 \tau+\gamma^{ij}S_{ij})\,,
\ee
\be{eq2}
(\partial_{t}-\beta^{j}\partial_{j})\phi=-\frac{1}{3} \phi \left(\alpha K-\partial_{k}\beta^{k}\right)\, , 
\ee
\be{eq3}
(\partial_{t}-\beta^{j}\partial_{j})\hat{\Gamma}^{i}=-2\tilde{A}^{ij}\partial_{j}\alpha+2\alpha\left[\tilde{\Gamma}^{i}_{kl}\tilde{A}^{kl}-3\tilde{A}^{ij}\frac{\partial_{j}\phi}{\phi} -\frac{2}{3}\tilde{\gamma}^{ij}\partial_{j}K  \right]
\ee
$$+2\tilde{\gamma}^{ki}\left( \alpha \partial_{k}\Theta - \Theta \partial_{k}\alpha -\frac{2}{3}\alpha K Z_{k}\right)
+2k_{3}\left(\frac{2}{3}\tilde{\gamma}^{ij}Z_{j}\partial_{k}\beta^{k}-\tilde{\gamma}^{jk}Z_{j}\partial_{k}\beta^{i} \right)$$
$$-\tilde{\Gamma}^{j}\partial_{j}\beta^{i}+\frac{2}{3}\tilde{\Gamma}^{i}\partial_{j}\beta^{j}+\frac{1}{3}\tilde{\gamma}^{ik}\partial_{j}\partial_{k}\beta^{j}+\tilde{\gamma}^{jk}\partial_{j}\partial_{k}\beta^{i}
-2 \alpha k_1 \tilde{\gamma}^{ij} Z_j
 -16\pi \alpha \tilde{\gamma}^{ik} S_{k}\,, $$
\be{eq4}
(\partial_{t}-\beta^{k}\partial_{k})\tilde{\gamma}_{ij}=-2\alpha \tilde{A}_{ij}+2\tilde{\gamma}_{k(i}\partial_{j)} \beta^{k} -\frac{2}{3}\tilde{\gamma}_{ij}\partial_{k} \beta^{k}\,,
\ee
\be{eq5}
(\partial_{t}-\beta^{k}\partial_{k})\tilde{A}_{ij}=e^{-4\pi \phi}\left[\alpha \tilde{R}^{\phi}_{ij}-\tilde{D}_{i}\tilde{D}_{j}\alpha+ \alpha D_{i}Z_{j}+ \alpha D_{j}Z_{i} \right]^{TF}
\ee
$$+\alpha \tilde{A}_{ij}(K-2\Theta)-2\alpha \tilde{A}_{ki}\tilde{A}_{j}^{k}+2\tilde{A}_{k(i} \partial_{j)} \beta^{k}-\frac{2}{3}\tilde{A}_{ij}\partial_{k} \beta^{k}-8\pi \alpha e^{-4\pi \phi}S_{ij}^{TF}\,,$$
\be{eq6}
(\partial_{t}-\beta^{j}\partial_{j})\Theta=\frac{1}{2}\alpha\left( R+2D_{i}Z^{i}-\tilde{A}_{ij}\tilde{A}^{ij}+\frac{2}{3}K^{2}-2\Theta K\right)-Z^{i}\partial_{i}\alpha -\alpha k_{1}(2+k_{2})\Theta -8\pi \alpha\, \tau\, . 
\ee
In this set of equations, $\alpha$ and $\beta^{i}$ stand respectively for the lapse and the shift functions of the ADM foliation, in term of which the unit four-vector $n^{\mu}$ normal to the space hyper-surface reads $n^{\mu}=(\alpha^{-1},-\beta^{i}\alpha^{-1})$ and $n_{\mu}=(-\alpha,0)$. We have denoted with $\tau$ the contraction to a scalar of the energy-momentum tensor with respect to the normal to the space hyper-surface, namely  $\tau=n_\mu \, n_\nu \, T^{\mu \nu}$, and with $S_i=n_\nu \, T^{\mu}_i$. We have also introduced the scalar functions $\tilde{A}$, $\tilde{D}$, $R$, $K$, $\tilde{\gamma}$ and $S$. The tensor $\tilde{\gamma}_{ij}\equiv e^{-4\phi}\gamma_{ij}$ is the conformal redefinition of  $\gamma_{ij}$, which is the spatial metric on the space-like hyper-surfaces and $\phi=\log({\rm det}(\gamma_{ij})/12)$ --- promoted to evolving variable --- while
\be{Atilde}
\tilde{A}_{ij}=e^{-4\pi \phi}\left(K_{ij}-\frac{1}{3}g_{ij}K  \right)\, , 
\ee
with $K$ the extrinsic curvature, and 
\be{Gammai}
\tilde{\Gamma}^{i}=\tilde{\gamma}^{jk}\tilde{\Gamma}_{jk}^{i}
\ee
are the Christoffel symbols of $\tilde{\gamma}_{ij}$. The tensor $S_{\mu\nu}=\gamma_{\mu}^{\alpha}\gamma_{\nu}^{\beta}T_{\alpha\beta}$ is the projection of energy-momentum tensor on the Cauchy surfaces. Finally, we have adopted the separation of variables $Z_{\mu}=(\Theta, Z_{i})$, where $\Theta\equiv Z_{0}$ and the conformal connection is substituted by $\hat{\Gamma}^{i}=\tilde{\Gamma}^{i}+2\tilde{\gamma}^{ij}Z_{j}$. 

One can consider the following 
gauge conditions: 
\be{g1}
\partial_{t}\alpha-\beta_{j}\partial^{j}\alpha=-2\alpha(K-2\Theta)\, , 
\ee
\be{g2}
\partial_{t}\beta^{i}-\beta_{j}\partial^{j}\beta^{i}=\frac{3}{4}B^{i}\, , 
\ee
\be{g3}
\partial_{t}B^{i}-\beta_{j}\partial^{j}B^{i}=\partial_{t}\hat{\Gamma}^{i}-\beta_{k}\partial^{k}\hat{\Gamma}^{i}-\eta B^{i}\, . 
\ee
As a convenient choice, we fix the free parameters $k_{i}$, as follows: $k_{1}=0.05$, $k_{2}=0$ and $k_{3}=1$ --- see {\it e.g.} Ref.~\cite{220}. 

In the next subsection, we will introduce the basic formulation of the gravitational wave signals used in our numerical simulations. 

\subsection{Gravitational wave signal}

From considering a generic binary source, 
one can numerical treat the emission of gravitational waves within the framework of the Newman-Penrose scalars \cite{Bishop:2016lgv}. The latter can be defined by contracting the Weyl tensor
\be{NP}
C_{\alpha\beta\mu\nu}=R_{\alpha\beta\mu\nu}-g_{\alpha[\mu}R_{\nu]\beta}+g_{\beta[\mu}R_{\nu]\alpha}+\frac{1}{3}g_{\alpha[\mu}g_{\nu]\beta}R
\ee
with an orthonormal null tetrad composed by the two pairs of normals $\vect{n}, \vect{l}$ and $\vect{m}, \vect{\bar{m}}$, such that  
\be{lnm}
l_{\mu}l^{\mu}=m_{\mu}m^{\mu}=n_{\mu}n^{\mu}=0\, ,
\ee
\be{lnm2}
m^{\mu}n_{\mu}=m^{\mu}l_{\mu}=0,\,\,\, l^{\mu}n_{\mu}=-1\,, \,\,\, m^{\mu} \bar{m}_{\mu}=1\,. 
\ee
In particular, the 
Newman-Penrose scalars are defined as follows: 
\be{zero}
\Psi_{0}=-C_{\alpha \beta \mu \nu} n^{\alpha} m^{\beta} n^{\mu} m^{\nu}\, , 
\ee
\be{uno}
\Psi_{1}=-C_{\alpha \beta \mu \nu} n^{\alpha} l^{\beta} n^{\mu} l^{\nu} \, , 
\ee
\be{due}
\Psi_{2}=-C_{\alpha \beta \mu \nu} n^{\alpha} m^{\beta} \bar{m}^{\mu} l^{\nu}\, , 
\ee
\be{tre}
\Psi_{3}=-C_{\alpha\beta\mu\nu} n^{\alpha} l^{\beta} \bar{m}^{\mu} l^{\nu}\, , 
\ee
\be{quattro}
\Psi_{4}=- C_{\alpha\beta\mu\nu} l^{\alpha} \bar{m}^{\beta} l^{\mu} \bar{m}^{\nu}\, .
\ee
In the asymptotic limit, $\Psi_{4}$ describes the gravitational radiation, namely 
\be{Psi4}
\Psi_{4}(r\rightarrow \infty)=\ddot{h}_{+}-i\ddot{h}_{X}\equiv \ddot{\bar{h}}\, . 
\ee

The Tetrad system used in {\it WeylScalar 4} of the {\it Einstein Toolkit}
--- as mentioned above, this is the open source code used for our numerical simulation ---
is specified by 
\be{lnm}
\vect{l}=\frac{1}{\sqrt{2}}(\vect{e}_{t}-\vect{e}_{R}),\,\,\,\ \vect{n}=\frac{1}{\sqrt{2}}(\vect{e}_{t}+\vect{e}_{R}),\,\,\,\vect{m}=\frac{1}{\sqrt{2}}(\vect{e}_{\theta}-i\vect{e}_{\phi})\, .
\ee

Eq.~\ref{Psi4} can be then conveniently expanded in spherical harmonics of weight $-2$,
\be{expansion}
\Psi_{4}(t,R,\theta,\phi)=\sum_{l=2}^{l_{max}}\sum_{m=-l}^{l}\Psi_{4,(-2)}^{lm}(t,R)Y^{lm}(\theta,\phi)\,,
\ee
as implemented in {\it Multipole} of {\it Einstein Toolkit}. 
Incidentally, we note that the {\it Peeling theorem} guarantees that the Newman-Penrose scalars, asymptotically in flat background, must fall down like $\Psi_{n}\sim r^{n-5}$, {\it i.e.} $\Psi_{4}\sim 1/r$. 

Integrating the $\Psi_{4}$ signal, the GW strain can be recovered to be 
\be{hlm}
\bar{h}_{lm}(t,r)=\int_{0}^{t}dt'\int dt'' \Psi_{4}^{lm}(t'',r)-C_{1}t-C_{0}\, , 
\ee
where 
\be{C1C0}
C_1=\Big|\frac{\partial \bar{h}}{\partial t}\Big|_{t=0},\,\,\,\, C_{0}=\bar{h}(t=0)\, . 
\ee

On the other hand, the asymptotic behaviour of the GW strain is simplified as in
\be{Psi}
h_{l,m}\Big|_{r\rightarrow \infty}(t)=\Psi_{l,m}(r,t)-\frac{l(l+1)}{2r}\int \Psi_{lm}(t,r)dt\,,
\ee
where $h_{l,m}\Big|_{r\rightarrow \infty}$ is the gravitational wave strain observed at infinity and $t$ is the retarded time. 

We may consider the evolution of the fourth Newman-Penrose scalar in the Kerr space-time. This can be connected to the Teukolsky's wave equation \cite{264}, namely 
\be{Kerr}
\left[\frac{(r^{2}+a^{2})^{2}}{\Delta}-a^{2}\sin^{2}\theta   \right]\frac{\partial^{2}}{\partial t^{2}}\Psi_{(-2)}
+4 \frac{a\, r\, M}{\Delta}\frac{\partial^{2}}{\partial t \partial \phi}\Psi_{(-2)}+\Big[\frac{a^{2}}{\Delta}-\frac{1}{\sin^{2}\theta}\frac{\partial^{2}}{\partial \phi^{2}}\Psi_{(-2)}  \Big]
\ee
 $$-\Delta^{2}\frac{\partial}{\partial r}\left(\frac{1}{\Delta}\frac{\partial}{\partial r}\Psi_{(-2)} \right)-\frac{1}{\sin \theta}
 \frac{\partial}{\partial \theta}\left(\sin \theta \frac{\partial}{\partial \theta}\Psi \right)
 +4\left[\frac{M(r^{2}-a^{2}}{\Delta}-r-i \,a\cos\theta   \right]\frac{\partial}{\partial}\Psi_{(-2)}
 $$
 $$+4\left[\frac{a(r-M)}{\Delta}+i\, {\rm cotg} \theta \right]\frac{\partial}{\partial \phi}\Psi_{-2}+2(2{\rm cotg}^{2}+1)\Psi_{(-2)}\,,$$
where $a=J/M$ is the Kerr spin parameter, $J$ is the total angular momentum, $M$ is the total mass of the system and $\Delta=r^{2}-2Mr+a^{2}$. We have introduced the definition of the Teukolsky's wave function $\Psi_{(-2)}$, which is in turn related to $\Psi_{4}$ by 
\be{PSIT}
\Psi_{(-2)}=(r-i\,a \,{\rm arcos}\theta)^{4}\Psi_{4}\, . 
\ee
The Fourier transform integral of the Teukolsky's wave function can be decomposed as follows: 
\be{Dec}
\Psi_{-2}=\int \sum_{l=0}^{l_{max}}\sum_{m=-l}^{l}\Psi_{(-2),lm,\omega}(r)S_{(-2),lm}^{\omega,a}(\theta,\phi)e^{-i \omega \phi}\, , 
\ee
where $S_{(-2),lm}$ is connected to the spherical harmonic functions as 
\be{SY}
S_{(-2)lm}^{\omega,a}=Y_{(-2)lm}+2a\omega\Big[-\frac{1}{l^{2}}\sqrt{\frac{(l+2)(l-2)(l+m)(l-m)}{(2l-1)(2l+1)}}Y_{(-2),m,l-1}
\ee
$$\times \frac{1}{(1+l)^{2}}\sqrt{\frac{(l+3)(l-1)(l+m-1)(l-m+1)}{(2l+1)(2l+3)}}    \Big]+O(a^{2}\omega^{2})\, . $$

From these expressions, one can find a relation of the wave function in terms of the strain 
\be{Psi}
\frac{1}{r^{3}}\Psi_{(-2),lm,\omega}=\int d\omega e^{-i\omega t}\Big[\Big( 1+\Big(-4\frac{ima}{l(1+l)}+i\frac{(l-1)(l+2)}{2\omega}\Big)\frac{1}{r}-\frac{1}{8}\frac{l(l-1)(l+1)(l+2)}{\omega^{2}}\frac{1}{r^{2}}\Big)h_{lm,\omega}
\ee
$$+2a\omega \Big(-\frac{1}{l^{2}}\sqrt{\frac{(l+2)(l-2)(l+m)(l-m)}{(2l-1)(2l+1)}}h_{l-1,m,\omega}$$
$$+\frac{1}{(l+1)^{2}}\sqrt{\frac{(l+3)(l-1)(l-m-1)(l-m+1)}{(2l+1)(2l+3)}}h_{l+1,m,\omega}\Big) \Big]+O\Big\{(a\omega)^{2},\frac{1}{(\omega r)^{3}}  \Big\}\, . $$

Finally, one can calculate the total radiated energy and the angular momentum from the gravitational strain 
\be{dEGW}
\dot{E}_{GW}=\frac{R^{2}}{16\pi}\int |\dot{h}(t,\theta,\phi)|^{2}\, ,
\ee
\be{JJ}
\dot{J}^{GW}=\frac{R^{2}}{16\pi}{\rm Re}\left[ \int d\omega\left(\partial_{\phi}\dot{\bar{h}}(t,\theta,\phi)  \right)h(t,\theta,\phi) \right]\, , 
\ee
where $R$ is the Black hole outermost radius. 

In the next subsection, we will introduce the main features specifying the Equation of State of the Mirror neutron stars, which are related to the precise spectrum of the gravitational waves emitted.

\subsection{EoS of Mirror Neutron stars}
\noindent 
A crucial point in order to calculate numerically the gravitational wave signals is modeling the equation of state (EoS) of Dark neutron stars. In a generic dark sector scenario, possible EoS of dark objects can
represent a serious issue, since in general may be completely different from the EoS of neutron stars. 
However, in this section we will argue that the Mirror symmetry can likely lead to the same EoS considered for ordinary neutron stars. This does not mean that Mirror neutron stars are, among possible dark stars, easier to be analyzed from numerical simulations. On the other hand, Mirror symmetry leads to the sharp prediction that the GW signal from Mirror neutron stars merging must have the same shape and features of the ordinary one. Ordinary neutron stars can be modeled within the framework of the Nambu--Jona-Lasino nuclear model.  The strong nuclear interactions among baryons are mediated by a tower of mesons:
the scalars $\sigma$, $\delta$; the pseudoscalars $\pi$, $K$, $\eta$, $\eta'$; the vectors $\rho$, $K^{*}$, $\omega$, $\phi$. The effective interactions among baryons and mesons read: 
\be{L}
\mathcal{L}_{s}=g_{s}\bar{B} B \Phi^{s}\, ,
\ee
\be{Lps}
\mathcal{L}_{ps}=g_{ps}\bar{B}i\gamma^{5}B \Phi^{ps}=g_{pv}\bar{B}\gamma^{5}\gamma^{\mu}B \partial_{\mu}\Phi^{ps}\, , 
\ee
\be{Lv}
\mathcal{L}_{v}=g_{v}\bar{B}\gamma^{\mu}B\Phi^{(v)\mu}+g_{t}\bar{B}\sigma^{\mu\nu}B(\partial_{\mu}\Phi_{\nu}^{(v)} -\partial_{\nu}\Phi_{\mu}^{(v)})\, . 
\ee
In Eqs.~\eqref{L}, \eqref{Lps} and \eqref{Lv}, $B$  denotes the baryon fields (spin 1/2), while $\Psi^{s,ps,v}$ are the corresponding scalar, pseudoscalar and vector fields. These Lagrangians are for iso-scalar mesons, while for iso-vector the equations should be modified according to $\Phi\rightarrow \vect{\tau}\cdot \vect{\Phi}$, where $\vect{\tau}$ are the isospin Pauli matrices. 

It is worth to note that the Mirror symmetry guarantee that every microscopic QCD couplings, as well as the non-perturbative QCD scale, must be equal to the ordinary one. So that, the effective Lagrangian for Mirror 
mesons and baryons --- we indicate here mirror baryons and mesons as $\tilde{B}$ and $\tilde{\Phi}$ --- must be 
\be{LM}
\mathcal{L}_{s}=g_{s}\widetilde{\bar{B}} \widetilde{B} \widetilde{\Phi}^{s}\, ,
\ee
\be{LpsM}
\mathcal{L}_{ps}=g_{ps}\widetilde{\bar{B}}i\gamma^{5}\widetilde{B} \widetilde{\Phi}^{ps}=g_{pv}\widetilde{\bar{B}}\gamma^{5}\gamma^{\mu}\tilde{B} \partial_{\mu}\tilde{\Phi}^{ps}\, , 
\ee
\be{LvM}
\mathcal{L}_{v}=g_{v}\widetilde{\bar{B}}\gamma^{\mu}\tilde{B}\tilde{\Phi}^{(v)\mu}+g_{t}\widetilde{\bar{B}}\sigma^{\mu\nu}\tilde{B}(\partial_{\mu}\tilde{\Phi}_{\nu}^{(v)} -\partial_{\nu}\tilde{\Phi}_{\mu}^{(v)})\, .
\ee
It is worth to remark that the couplings of Eqs.(\ref{LM}), (\ref{LpsM}) and (\ref{LvM}) are exactly the same as in Eqs.(\ref{L}), (\ref{Lps}) and (\ref{Lv}), as well as the mesons and baryons masses \footnote{In principle, possible mixing terms among neutral ordinary baryons and mesons and their mirror twins can be introduced. However, the large chemical potential of the neutron star will completely suppress these possible mixing terms. }. \\

The one-meson exchange potential in baryon-baryon scattering casts 
\be{VVV}
\langle p_{1}'p_{2}'|V_{\Phi}|p_{1}p_{2}\rangle=\frac{\bar{u}(p_{1}')g_{\Phi_{1}}\Gamma_{\Phi}^{(1)}u(p_{1})P_{\Phi} \bar{u}(p_{2}')g_{\Phi_{2}}\Gamma_{\Phi}^{(2)} u(p_{2})}{(p_{1}-p_{1}')^2-m_{\Phi}^{2}}\,.
\ee
In  Eq.(\ref{VVV}), $p_{1,2}$ and $p_{1,2}'$ are the four-momenta of the in and out scattering baryons. The tensorial structure of the (pseudo)scalar and vector propagators read respectively $P_{\Phi}^{s,ps}=1$ and, for a choice of the signature $g_{\mu\nu}=(1,-1,-1,-1)$, 
$$P_{\Phi}^{v}=P_{\mu\nu}=-g_{\mu\nu}+\frac{q_{\mu}q_{\nu}}{m_{\Phi}^{2}}\,.$$
The baryon spinors are denoted by $u$, while $\Gamma_{\Phi}$ represent 
$$\Gamma_{s}=1\!\!1,\,\,\,\, \Gamma_{ps}=i\gamma^{5},\,\,\,\Gamma_{v}=\gamma^{\mu},\,\,\, \Gamma_{t}=\sigma_{\mu\nu}\, . $$
Finally, with $q=p_{1}-p'_{1}$ we denote the four momentum transferred, while $m_{\Phi}$ is the mirror meson mediator mass.

Consequently, the non-relativistic effective Mirror baryon - Mirror baryon potential, in local approximation,
is exactly the same as the ordinary one, {\it i.e.}
\be{Pot}
V(\vect{r})=\sum_{\Phi}\Big\{C_{C_{\Phi}}+C_{\sigma_{\Phi}}\vec{\sigma}_{1}\cdot \vec{\sigma}_{2}+C_{LS_{\Phi}}\left(\frac{1}{m_{\Phi}r}+\frac{1}{(m_{\Phi}r)^{2}} \right) \vec{L} \cdot \vec{S}
\ee
$$+\, C_{T_{\Phi}}\left( 1+\frac{3}{m_{\Phi}r}+\frac{3}{(m_{\Phi}r)^{2}}\right)\vec{S}_{12}\Big\}\frac{e^{-m_{\Phi}}}{r}\, , $$
In Eq.~\eqref{Pot} the $C$s are numerical combinations of the coupling constants and the mirror baryon masses. The total angular momentum and the total spin of the system are denoted respectively with $\vec{L}$ and $\vec{S}$, while 
\be{S12}
\vec{S}_{12}(\hat{r})=2(\sigma_{1} \cdot \hat{r})(\sigma_{2} \cdot \hat{r})-(\sigma_{1}\cdot \sigma_{2})\, ,
\ee
with $\hat{r}=\vect{r}/|r|$. 

In every interaction vertices, one must also include the form factors, which can be modeled in two possible but mutually consistent semi-analytical ways, {\it i.e.}  
\be{F}
F_{\alpha}=\left(\frac{\Lambda_{\alpha}^{2}-m_{\alpha}^{2}}{\Lambda_{\alpha}^{2}+|\vec{k}|^{2}}\right)^{n_{\alpha}}
\ee
or 
\be{FF}
F_{\alpha}={\rm exp}\left( -\frac{|\vec{k}|^{2}}{2\Lambda_{\alpha}}\right)\,.
\ee
In \eqref{F} and \eqref{FF}, $\vec{k}$ is the 3-momentum transferred, $\Lambda_{\alpha}\sim 1.2 \div 2\, {\rm GeV}$ is the cut-off mass scale, while $n_{\alpha}$ is a constant adimensional parameter which is usually consider as $=1$ or $=2$ (depending on the choice of the $\Lambda_{\alpha}$ scales). The form factors cut-off the divergences of the effective baryon-meson model. Since they are completely determined by microscopic couplings and the $\Lambda_{QCD}$ scale, Mirror baryons and mesons will have the same form factors than ordinary twin particles. \\

The Mirror pseudo-scalar mesons interaction with Mirror baryons can be described as
\be{LBB}
\mathcal{L}=\langle i \widetilde{\bar{B}}\gamma^{\mu}D_{\mu}\widetilde{B}-M_{0}\widetilde{\bar{B}}\widetilde{B}+\frac{f_{1}}{2}\widetilde{\bar{B}}\gamma^{\mu}\{\widetilde{u}_{\mu},\widetilde{B} \}+\frac{f_{2}}{2}\widetilde{\bar{B}}\gamma^{\mu}\gamma^{5}[\widetilde{u}_{\mu},\widetilde{B}]\rangle\,,
\ee
where 
\begin{align}\label{matrix'1}
\widetilde{B}=\left( \begin{array}{ccc} \frac{1}{\sqrt{2}}\widetilde{\Sigma}^{0}+\frac{1}{\sqrt{2}}\widetilde{\Lambda}  & \widetilde{\Sigma}^{+} & \widetilde{p} \\ 
\widetilde{\Sigma}^{-} & -\frac{1}{\sqrt{2}}\widetilde{\Sigma}^{0}+\frac{1}{\sqrt{6}}\widetilde{\Lambda} & \widetilde{n} \\
- \widetilde{\Xi}^{-} & \widetilde{\Xi}^{0} & -\frac{2}{\sqrt{6}}\widetilde{\Lambda} \end{array} \right),
\end{align}
$D_{\mu}=\partial_{\mu}+[\Gamma_{\mu}, ... ]$ is the covariant derivative, $M_{0}$ is the Mirror octet baryon mass --- which is equal to the ordinary one --- in the chiral limit; the brackets denote the trace over the $SU(3)$-flavor symmetry; $f_{1,2}$ are exactly the same as ordinary matter, and furthermore can be constrained by laboratory experiments to be $f_{1}+f_{2}=g_{A}\simeq 1.26$ ($f_{1}=0.80$, $f_{2}=0.26$ in order to fit semi-leptonic decays $B\rightarrow B'+e+\bar{\nu}_{e}$), while  $g_{A}$ is the axial-vector coupling; 
$\widetilde{u}_{\mu}$ stands for $i\partial_\mu \widetilde{U}$, in turns equal to 
\be{UUU}
%u_{\mu}=i u^\dagger \partial_{\mu} U u^{\dagger} ,\,\,\, 
\widetilde{U}=\exp\left(\frac{2i\widetilde{P}}{\sqrt{2}F_{\pi}} \right)\, , 
\ee
\begin{align}\label{matrix'1}
\widetilde{P}=\left( \begin{array}{ccc} \frac{1}{\sqrt{2}}\widetilde{\pi}^{0}+\frac{1}{\sqrt{2}}\widetilde{\eta} & \widetilde{\pi}^{+} & \widetilde{K}^{+} \\ 
\widetilde{\pi}^{-} & -\frac{1}{\sqrt{2}}\widetilde{\pi}^{0}+\frac{1}{\sqrt{6}}\widetilde{\eta} & \widetilde{K}^{0} \\
- \widetilde{K}^{-} & \widetilde{\bar{K}}^{0} & -\frac{2}{\sqrt{6}}\widetilde{\eta}
\end{array} \right),
\end{align}
where $F_{\pi}=92.4\, {\rm MeV}$ is the weak Mirror pion decay constant, 
which is equal to the ordinary one. 

Thus, since the nuclear physics of the Mirror neutron stars is exactly the same as ordinary ones, we can safely consider the same way of modeling the EoS for Mirror neutron stars. In particular, we consider the four standard EoS models: {\bf SLy} \cite{T142}, {\bf AP4} \cite{T140}, {\bf H4} \cite{T101} and {\bf MS1} \cite{T143}.\\

The {\bf SLy} model has a simple semi-analytical representation in terms of 18th fit parameters: 
\be{zeta}
\zeta=\frac{a_{1}+a_{2}\eta+a_{3}\eta^{3}}{1+a_{4}\eta}f(a_{5}(\eta-a_{6}))
+(a_{7}+a_{8}\zeta)f(a_{9}(a_{10}-\zeta))
\ee
$$+(a_{11}+a_{12}\eta)f(a_{13}(a_{14}-\eta))+(a_{15}+a_{16}\eta)f(a_{17}(a_{18}-\eta))\, , $$
where 
\be{zetaepsolon}
\zeta=\log(P^2/{\rm dyn\, cm^{-2}}),\,\,\,\ \zeta=\log(\rho/{\rm g\, cm^{-3}}),\,\,\, f(x)=1/(1+e^{x})\, , 
\ee
$a_{1,...,18}$ are the fit parameters defined in Ref.~\cite{T142}. \\

The {\bf AP4} as a semi-analytical representation defined as follows: 
\be{zeta}
\zeta=\zeta_{low}f(a_{1}(\eta-c_{11}))+f_{0}(a_{2}(c_{12}-\eta))\zeta_{high}\,,
\ee
where
\be{zetalow}
\zeta_{low}=[c_{1}+c_{2}(\eta-c_{3})^{c_{4}}]f(c_{5}(\zeta-c_{6}))+(c_{7}+c_{8}\eta)f(c_{9}(c_{10}-\eta))\, , 
\ee
\be{zetahigh}
\zeta_{high}=(a_{3}+a_{4}\zeta)f(a_{5}(a_{6}-\eta))+(a_{7}+a_{8}\eta+a_{9}\eta^{2})f(a_{10}(a_{11}-\eta))\, . 
\ee
Here the fit parameters are 
$$c_{1}=10.6557,\,\,\,c_{2}=3.7863,\,\,\,c_{3}=0.8124,\,\,\,c_{4}=0.6823\, , $$
$$c_{5}=3.5279,\,\,\, c_{6}=11.8100,\,\,\, c_{7}=12.0584,\,\,\, c_{8}=1.4663\, , $$
$$c_{9}=3.4952,\,\,\, c_{10}=11.8007,\,\,\, c_{11}=14.4114,\,\,\, c_{12}=14.4081\, , $$

$$a_{1}=4.3290,\,\,\, a_{2}=4.3622,\,\,\, a_{3}=9.1131,\,\,\, a_{4}=-0.4751\, , $$
$$a_{5}=3.4614,\,\,\, a_{6}=14.8800,\,\,\, a_{7}=21.3141,\,\,\, a_{8}=0.1023\, , $$
$$a_{9}=0.0495,\,\,\, a_{10}=4.9401,\,\,\, a_{11}=10.2957\, . $$
\\

The {\bf H4} and {\bf MS1} have the same low and high density EoS. The low density profile can be modeled as 
\be{zeta}
\zeta=\Gamma_{i}K_{i}\eta
\ee
in each density interval $[\rho_{i},\rho_{i+1}]$. Furthermore, both {\bf H4}, {\bf MS1} can be modeled with the following four sets of parameters: 
$$\Gamma_{0}=1.584,\,\,\ K_{0}=6.801\times 10^{-11}\, ,$$
$$\Gamma_{1}=1.287,\,\,\, K_{1}=1.062\times 10^{-6}\, , $$
$$\Gamma_{2}=0.6224,\,\,\, K_{2}=50.327\, ,$$
$$\Gamma_{3}=1.357,\,\,\, K_{3}=3.999\times 10^{-8}\, . $$
\\

On the other hand, within the high density region of the neutron star, {\bf H4} and {\bf MS1} have the different parameterizations: 

{\bf H4:}
$$\Gamma_{4}=2.909,\,\,\, \Gamma_{5}=2.246,\,\,\, \Gamma_{6}=2.144\, , $$
with a constant density core 
$$e^{\eta}\equiv \rho[g/cm^{3}]=0.888 \times 10^{14}\, ;$$

{\bf MS1:}
$$\Gamma_{4}=3.224,\,\,\, \Gamma_{5}=3.033,\,\,\, \Gamma_{6}=1.325\, ,$$
with 
$$e^{\eta}\equiv \rho[g/cm^3]=0.942 \times 10^{14} \, .$$

%Read J S, Lackey B D, Owen B J and Friedman J L 2009 Physical Review D 79 124032 ISSN 1550-7998 URL http://link.aps.org/doi/10.1103/PhysRevD.79.124032

In Fig.1, the EoS belonging to the models used for our numerical simulations are displayed,
while in Fig.2, the corresponding GW signals, calculated within the framework of the Newman-Penrose formalism, are compared.

\section{Conclusions and remarks} \label{co}

\noindent 
We discussed the intriguing possibility that the merging of exotic Dark Stars, not predicted by the Standard Model of particle physics and General Relativity, can be tested thanks to the detection of gravitational waves signals that show no correlation with electromagnetic signals. This is a strategy that is allowed now days by the recent experimental observations of correlated gravitational and electromagnetic signals, arisen out of the LIGO/VIRGO and FERMI/INTEGRAL collaborations. 

We discussed this possibility examining signals consistently produced by ``invisible'' candidates of Dark Matter, {\it i.e.} non interacting with the electromagnetic sector. Specifically, we focused on the framework of Mirror Dark Matter, a long-standing candidate for Dark Matter that predicts similar microphysics laws than the ordinary Standard Model of particle physics, while doubling entirely the gauge sector of the latter. \\

Specifically, to claim detection of Mirror Dark Matter via tests of Mirror Stars Merging we must observe gravitational waves signals with {\it exactly} the same main features as the ones predicted by ordinary matter, and at the same time correlation to electromagnetic signals proper of ordinary Stars Merging must be rejected as hypothesis. Indeed the equations of state of Mirror Neutron Stars are exactly the same as in the ordinary sector. This is a sharp prediction arising from Mirror Symmetry, and constitutes a fundamental principle of Nature within the Mirror paradigm. 

Thanks to this, despite of the astrophysical complexity of the Mirror Dark Halos and their highly unpredictable hidden structure formation, numerical investigations already carried out for Ordinary Neutron Stars can be deployed in the search for Mirror Neutron Stars merging, in the gravitational channel. In this scenario, there is a fundamental {\it caveat} to take care of. Mirror systems must be totally invisible in the electromagnetic channel, {\it i.e.} the hypothesis of correlations to electromagnetic signals that are proper of ordinary stars merging must be rejected. This highly motivates a multi-messenger approach in order to discriminate signals arising from the merging of {\it Exotic} Stars, from signals predicted by the microphysics (Standard Model of particle physics and General Relativity) of Ordinary systems.\\
 
We focused on the case of Mirror Neutron stars, but of course {\it Mirror Supernovae} would also represent valid candidates to test signals of gravitational waves originated by the Mirror sector. Remarkably, a third channel would be present in this latter case to corroborate detection of the event. This further channel is represented by the detection of neutrinos signals, to be combined now with gravitational waves signals. Correlations between the gravitational waves signals and the neutrinos signals allows to identify the class of objects that are merging. The presence of neutrinos channel in Mirror Supernovae is due to the fact that Mirror Dark Matter is compatible with large --- enough to be observed at cosmological distances --- mixing among Mirror neutrinos and ordinary neutrinos \cite{Berezhiani:1995yi,Foot:1995pa}.\\

It is worth to remark that the observation of a lack of correlation among a gravitational wave signal and an electromagnetic counterpart cannot be considered as a direct detection observation of Mirror dark matter. 
Tthe hypothetical detection of such a signal, without an electromagnetic counterpart, would necessary request to revisit standard astrophysics: such an observation could not be explained by Ordinary neutron stars (no electromagnetic counterpart) and by black holes (the gravitational signal waveform is completely different). In this sense, the hypothetical observation that we envisage in this paper should be interpreted as an indirect "hint", favoring the existence of exotic dark stars, such as Mirror Neutron stars. On the other hand, the no-observation result would be important in order to exclude dark sector scenarios 
predicting a high probability of dark compact objects merging. 

Another interesting possibility is that Mirror Dark Matter coexists with axion dark matter. The axion field can be introduced as a common Peccei-Quinn shift symmetry among the Ordinary and the Mirror sector. In this  picture, the Milky Way Mirror halo can be distributed as a Silk Disk, {\it i.e.} a Double Disk Dark Matter scenario can be envisaged. On the other hand, axion dark matter can be mostly distributed outside the Milky way disk, organized as an elliptical halo. In this case, it is also conceivable the existence of exotic axion stars due to the overdensity instabilities of the axion condensate. In this scenario, the merging of boson stars can predict a peculiarly interesting gravitational wave signal --- see {\it e.g.} Refs. \cite{Dev:2016hxv,Palenzuela:2017kcg} for recent numerical analyses on Boson Stars merging. According to this hypothesis, observations of gravitational waves signal from Mirror neutron stars merging are predicted in the Milky Way plane, while axion stars merging are expected from the sky regions outside the Milky Way plane. 

To conclude, we emphasize that the strategy discussed in this paper, which mainly pertains the theoretical framework of Mirror Dark Matter, can be opportunely generalized so encode several other dark gauge sectors, as for instance:
\begin{itemize}  
\item
Minimal instantiation of Mirror Dark Matter (discrete symmetries scenarios, IQCD, extra SU(2), etc...);
\item
Dark Silk Disk;
\item
Technicolor. 
\end{itemize}

\vspace{1cm} 

{\large \bf Acknowledgments} 
\vspace{4mm}

\noindent
We thank Nico Junes for useful discussions and remarks on the aspects about
numerical relativity and GW observational.  
A.A. wishes to acknowledge Zurab Berezhiani for many interesting discussions on these subjects.  A.M. wishes to acknowledge Stephon Alexander for inspiring discussions on the relevant topics of the paper, and support by the Shanghai Municipality, through the grant No. KBH1512299, and by Fudan University, through the grant No. JJH1512105. 

\section*{Appendix: more about the structure formation problem}

In this section, we will add further considerations and remarks on the insidious problem that concerns the structure formation in Mirror dark matter halos. 

The main open problem is how Mirror Dark Matter (MDM) can form an extended dark halo, which is dissipative. One might expect indeed that Mirror Matter forms Galactic Silk Disks despite of extended halos. The Milky Way (MW) disk predicted by the Mirror sector must have different radius and thickness. In principle, this is not incompatible with local Dark Matter halo constraints. As it is well known, the total matter surface density at the Sun radius is (68$\pm$ 4) $M_{\odot}/pc^{2}$ --- see {\it e.g.} Ref.\cite{87}. Since, the amount of visibile energy density contributes to the total amount as (38$\pm$ 4) $M_{\odot}/pc^{2}$ \cite{87}, the surface density of the Mirror matter density should account for the remaining part, compatible with the presence of a Mirror Silk Disk. A Dark Matter Silk Disk cannot explain the MW galaxy rotational curves without considering other Dark Matter components, such as axions or other parallel asymmetric Mirror sectors. Such a multi-component Dark Matter scenario seems to lie into the class of models that are dubbed {\it Double Disk Dark Matter} and were recently analyzed in Refs.~\cite{Fan:2013tia,Fan:2013yva,McCullough:2013jma,Kramer:2016dew,Kramer:2016dqu}.\\

Nonetheless these considerations cannot be claimed to be definitive, since any numerical N-body simulations within the framework of MDM has been yet performed. For example, it is still possible that the high rate of supernovae explosions in the Mirror halo core could  cause a back-reaction effect that is able to reheat Mirror nuclei and can then compete with dissipative contraction. \\

Another possibility that is still open, also suggested in Ref.~\cite{Cerulli:2017jzz}, is that the Mirror stars formation is so rapid --- generally, it is faster than stars formation in the Ordinary World --- that stars arise faster than the Silk Disk would originate from dissipative cooling. In this latter case, the Mirror halo that emerges from the dissipative process is similar to an elliptical galaxy rather than a Silk Disk. Such a possibility may be suggested also following the Jeans criteria \cite{Pad}. The Jeans mass is indeed smaller within the Mirror matter framework. Mirror matter is cooler and helium dominated, {\it i.e.} the star formation is more efficient.  Let us remind that, from present data, the most accreditate mechanism for the formation of the Elliptical galaxies in our Ordinary Matter sector is imagined to be the merging of galactic spiral galaxies. Nonetheless, our galactic dark halo may be formed too from the merging of two Silk Disks, generating an elliptical halo \cite{elliptical}. It is not clear though how in this case the MW ordinary disk would be preserved from such a collision. Still, it is possible that our MW Dark halo is formed from the merging of a almost totally dominated elliptical Mirror galaxy and a Double Disk system. \\

From these considerations, we may argue that the internal structure of the MW Dark halo is the least predictable issue of MDM, as it deserves sophisticated numerical N-body simulations. However current astrophysical observations, {\it e.g.} micro-lensing constraints from Massive Compact Astrophysical Halo objects and Gravitational Waves emissions, allow to limit {\it a posteriori} available theoretical scenarios. For example, MACHO and EROS collaborations set sever limits on the ratio of Massive Compact Astrophysical Halo Objects (MACHOs). For instance, they exclude that the halo has more than the $0.1\%$ of MACHOs or so, within a mass spectrum of approximately $10^{-6}\div 10\, M_{\odot}$ \cite{Alcock:2000ph,Tisserand:2006zx}. The hypothesis that the Mirror MW halo could be an elliptical galaxy seems to be disfavored at a first sight. Conversely, the Double Dark Disks hypothesis can be subtly compatible with MACHO/EROS analysis, since the measure is performed pointing toward the Large Magellanic Cloud, outside the direction of the galactic plane. In this case, the presence of MACHOs on the Dark Silk Disk is still unconstrained and the hypothesis cannot be excluded. \\

It is worth noting that even this conclusion might not be completely correct. This is because limits arisen from the MACHO and EROS collaborations are highly model dependent, relying on the choice of the Dark Halo models. The latter are assumed by the MACHO and EROS collaborations to be deformations of isothermal-like halos. Nonetheless, constraints from MACHO/EROS can be further relaxed, if one assumes a co-rotating or counter-rotating Mirror halo. The optical depth factor, from which the average MACHO masses and ratio are related, is quadratically dependent on the relative velocity among the Earth and the MACHO objects. Then, a co-rotating halo will change the relative velocity, making the MACHOs detection more elusive.  Since Elliptical galaxies are observed to have a rotation of almost $50\div 100\, {\rm km/s}$, this can radically change the analysis. For instance, comparing the assumptions for the Mirror MW halo with the ones for the MACHO/EROS halo models, it is straightforward to note that the estimate of the ratio may change by a factor $2$ or even more \cite{Zurab}.\\
 
Analogue comments concern constraints on the Mirror Dark halo that arise from the Bullet Cluster observations \cite{Markevitch:2003at}. The Bullet Cluster imposes strong limits on the Dark Matter self-collisional cross section $\sigma_{s.c.}$ over the dark matter particle mass $m$. The bound on the ratio, which is approximately $\sigma_{s.c.}/m<10^{-23}\, {\rm cm^{2}/GeV}$, seems to be a {\it gravestone} for Mirror matter appearing in the form of a plasma of collisional nuclei. Nonetheless, if Mirror Dark halos of Bullet Clusters are mostly organized in mirror stars, the Mirror star-star collision probability over the star mass is highly suppressed, and again Mirror Dark Matter can be subtly compatible with Clusters data. Indeed, contrary to the WIMP paradigm, Mirror Halo can be organized in several different structures, including plasma like forms, very similarly to what observed in the ordinary sector. Such a rich cosmological picture offers also the possibility to explain possible future hints of dark matter self-collisionality from specific clusters.

\appendix

%\section{A brief review of the AGK rules }

\end{document}